\documentclass[sigconf,preprint]{acmart}

\usepackage{booktabs} 
\usepackage[utf8]{inputenc}
\usepackage{verbatim}
\usepackage{algorithm}
\usepackage[noend]{algpseudocode}

\newcommand{\throw}{\Uparrow}
\newcommand{\annotate}{annotate}
\newcommand{\labexp}{labexp}
\newcommand{\addlab}{addlab}
\newcommand{\calck}{calck}
\newcommand{\fst}{FIRST}
\newcommand{\flw}{FOLLOW}
\newcommand*\Let[2]{\State #1 $\gets$ #2}
\newcommand{\Epsi}{\varepsilon}
\newcommand{\Rp}{R^\prime}

\setcopyright{rightsretained}

\acmDOI{10.1145/3264637.3264638}

\acmISBN{978-1-4503-6480-5}

\acmConference[SBLP'18]{Brazilian Symposium on Programming Languages}{September 2018}{São Carlos, São Paulo BR}
\acmYear{2018}
\copyrightyear{2018}

\acmPrice{15.00}

\begin{document}
\title{Towards Automatic Error Recovery in Parsing Expression Grammars}

\author{S{\'e}rgio Queiroz de Medeiros}
\affiliation{%
  \institution{Federal University of Rio Grande do Norte}
  \country{Brazil}
}
\email{sergiomedeiros@ect.ufrn.br}

\author{Fabio Mascarenhas}
\affiliation{%
  \institution{Federal University of Rio de Janeiro}
  \country{Brazil}
}
\email{fabiom@dcc.ufrj.br}


\begin{abstract}
Error recovery is an essential feature for a parser that
should be plugged in Integrated Development Environments (IDEs),
which must build Abstract Syntax Trees (ASTs) even for syntactically
invalid programs in order to offer features such as automated refactoring and code
completion.

Parsing Expressions Grammars (PEGs) are a formalism that
naturally describes recursive top-down parsers using a restricted form of
backtracking. Labeled failures are a
conservative extension of PEGs that adds an error reporting
mechanism for PEG parsers, and these labels can also be associated
with recovery expressions to also be an error recovery mechanism.
These expressions can use the full expressivity of PEGs to recover
from syntactic errors.

Manually annotating a large grammar with labels and recovery expressions
can be difficult. In this work, we present an algorithm that automatically
annotates a PEG
with labels, and builds their corresponding recovery expressions.
We evaluate this algorithm by adding error recovery to the parser
of the Titan programming language. The results shown that
with a small amount of manual intervention our algorithm can
be used to produce error recovering parsers for PEGs where
most of the alternatives are disjoint.
\end{abstract}

%
%



\maketitle

\section{Introduction}
\label{sec:intro}

Integrated Development Environments (IDEs) often require parsers
that can recover from syntax errors and build syntax trees
even for syntactically invalid programs, in other to conduct
further analyses necessary for IDE features such as automated refactoring
and code completion.

Parsing Expression Grammars (PEGs)~\cite{ford2004peg} are a formalism
used to describe the syntax of programming languages, as an alternative
for Context-Free Grammars (CFGs).
We can view a PEG as a formal description of a recursive top-down parser
for the language it describes. PEGs have a concrete syntax based on
the syntax of {\em regexes}, or extended regular expressions.
Unlike CFGs,
PEGs avoid ambiguities in the definition of the grammar's
language by construction, due to the use of an {\em ordered choice} operator.

The ordered choice operator naturally maps to restricted (or local)
backtracking in a recursive top-down parser.
The alternatives of a choice are tried in
order; when the first alternative recognizes an input prefix,
no other alternative of this choice is tried, but when an
alternative fails to recognize an input prefix, the parser
backtracks to the same input position it was before
trying this alternative and then tries the next one.

A naive interpretation of PEGs is problematic when dealing with
inputs with syntactic errors, as a failure during parsing an input
is not necessarily an error, but can be just an indication that the
parser should backtrack and try another alternative. 
Labeled failures~\cite{maidl2013peglabel,maidl2016peglabel} are
a conservative extension of PEGs that address this problem of
error reporting in PEGs by using explicit error labels, which are distinct
from a regular failure. We throw a label to signal an error
during parsing, and each label can then be tied to an
specific error message.

We can leverage the same labels to add an error recovery mechanism,
by attaching a recovery expression to each label. This expression
is just a regular parsing expression, and it usually skips the erroneous
input until reaching a synchronization point, while producing a
dummy AST node of the type that was expected.~\cite{medeiros2018sac}.

Labeled failures produce good error messages and error recovery,
but they can add a considerable annotation burden in large grammars,
as each point where we want to signal and recover from a syntactic error
must be explicitly marked.

In this paper, we present an algorithm that automatically annotates a PEG
with labels, and builds their corresponding recovery expressions.
We also evaluate the use of this algorithm to build an error recovering
parser for the Titan programming language. The results show that
with a small amount of manual intervention our algorithm can
be used to produce an error recovering parser from a PEG where
most of the alternatives are disjoint.

The remainder of this paper is organized as follows: 
Section~\ref{sec:pegs} discusses error recovery in PEGs
using labeled failures and recovery expressions;
Section~\ref{sec:algo} presents our algorithm to automatically
annotate a PEG with labels and to associate a recovery expression
to each label;
Section~\ref{sec:eval} evaluates the use of our algorithm
to annotate the grammar of the Titan programming language;
Section~\ref{sec:rel} discusses related work on error reporting and error recovery;
finally, Section~\ref{sec:conc} gives some concluding remarks.

\section{Error Recovery in PEGs with Labeled Failures}
\label{sec:pegs}

\begin{figure}[t]
{\small
\centering
\begin{align*}
\it{Prog} & \leftarrow {\tt PUBLIC\;CLASS\;NAME\;LCUR\;PUBLIC\;STATIC\;VOID\;MAIN}\\
& \;\;\;\;\;{\tt LPAR\;STRING\;LBRA\;RBRA\;NAME\;RPAR}\;\it{BlockStmt}\;{\tt RCUR}\\
{\it BlockStmt} & \leftarrow {\tt LCUR}\;{\it (Stmt)*}\;{\tt RCUR}\\
{\it Stmt} & \leftarrow {\it  IfStmt\;/\;WhileStmt\;/\; PrintStmt\;/\;DecStmt\;/\;AssignStmt\;/}\\
& \;\;\;\;\;{\it BlockStmt}\\
{\it IfStmt} & \leftarrow {\tt IF\;LPAR}\;{\it Exp}\;{\tt RPAR}\;{\it Stmt}\;{\it (}{\tt ELSE}\;{\it Stmt\;/\;\varepsilon)}\\
{\it WhileStmt} & \leftarrow {\tt WHILE\;LPAR}\;{\it Exp}\;{\tt RPAR}\;{\it Stmt}\\
{\it DecStmt} & \leftarrow {\tt INT\;NAME}\;({\tt  ASSIGN}\;{\it Exp}\;/\;\varepsilon)\;{\tt SEMI}\\
{\it AssignStmt} & \leftarrow {\tt NAME\;ASSIGN}\;{\it Exp}\;{\tt SEMI}\\
{\it PrintStmt} & \leftarrow {\tt PRINTLN\;LPAR}\;{\it Exp}\;{\tt RPAR\;SEMI}\\
\it{Exp} & \leftarrow \it{RelExp \; ({\tt EQ} \; RelExp)*} \\
\it{RelExp} & \leftarrow \it{AddExp \; ({\tt LT} \; AddExp)*} \\
\it{AddExp} & \leftarrow \it{MulExp \; (({\tt PLUS} \; / \; {\tt MINUS}) \; MulExp)*} \\
\it{MulExp} & \leftarrow \it{AtomExp \; (({\tt TIMES} \; / \; {\tt DIV}) \; AtomExp)*} \\
\it{AtomExp} & \leftarrow \it{{\tt LPAR} \; Exp \; {\tt RPAR} \; / \; {\tt NUMBER} \; / \; {\tt NAME}}
\end{align*}
}
\vspace{-0.7cm}
\caption{A PEG for a tiny subset of Java}
\label{fig:javagrammar}
\end{figure}

In this section we present a short introduction to labeled
PEGs and discuss how to build an error recovery mechanism for PEGs
by attaching a recovery expression to each labeled failure.
A more detailed presentation of labeled PEGs, which
includes its formal semantics, can be found
in our previous work~\cite{medeiros2018sac}.

A labeled PEG $G$ is a tuple $(V,T,P,L,R,{\tt fail}, p_{S})$ where
$V$ is a finite set of non-terminals, $T$ is a finite set of terminals,
$P$ is a total function from non-terminals to \emph{parsing expressions},
$L$ is a finite set of labels, $R$ is a function from labels to
\emph{parsing expressions}, ${\tt fail} \notin L$ is a failure label,
and $p_{S}$ is the initial parsing expression.
We will use the term {\em recovery expression} when referring to
the parsing expression associated with a given label.

We describe the function $P$ as a set of rules of the form
$A \leftarrow p$, where $A \in V$ and $p$ is a parsing
expression.
A parsing expression, when applied to an input string, either
produces a label, associated with an input position, or consumes a prefix of the input and returns the
remaining suffix. If the expression produces {\tt fail} we say that it failed.
The abstract syntax of parsing expressions is
given as follows, where $a$ is a terminal, $A$ is a non-terminal,
and $p$, $p_1$ and $p_2$ are parsing expressions, and $l$ is
a failure label:
\[\it{
p = \varepsilon \; | \; a \; | \; A \; | \; p_1 p_2 \; | \;
  p_1 / p_2 \; | \; p\!* \; | \; !p \; | \; \throw^l
}\]

Informally,
$\varepsilon$ successfully matches while not consuming any input;
$a$ matches and consumes itself or fails otherwise;
$A$ tries to match the expression $P(A)$;
$p_1 p_2$ tries to match $p_1$ followed by $p_2$;
$p_1 / p_2$ tries to match $p_1$;
if $p_1$ fails, i.e., the result of matching $p_1$
is {\tt fail}, we try to match $p_2$;
$p*$ repeatedly matches $p$ until $p$ fails, that is, it
consumes as much as it can from the input;
$!p$ succeeds if the input does not match $p$ producing any label,
and fails when the input matches $p$, not consuming any input in either case;
we call it the negative predicate or the lookahead predicate;
$\throw^{l}$, where $l \in L$, generates a failure
with label $l$, and in case $l$ has an associated recovery expression
it will be used to match the input from the point where $l$ was
thrown.

A label $l \neq {\tt fail}$ thrown by $\throw$ cannot
be caught by an ordered choice or a repetition,
so it indicates an actual error during parsing, while {\tt fail} 
indicates that the parser should backtrack. 
The lookahead operator $!$ captures any label and turns it into a success,
while turning a success into a {\tt fail} label. The rationale is
that errors inside a syntactic predicate are expected and not actually
syntactic errors in the input.

Figure~\ref{fig:javagrammar} shows a PEG for a tiny subset of Java, where
lexical rules (shown in uppercase) have been elided. 
While simple (this PEG is equivalent to an LL(1) CFG),
this subset is a good starting point to discuss error recovery
in the context of PEGs.

To get a parser with error recovery, we first need to have
a parser that correctly reports errors. One popular error
reporting approach for PEGs is to report the farthest
failure position~\cite{ford2002packrat,maidl2016peglabel},
an approach that is supported by PEGs with
labels~\cite{medeiros2018sac}. However, the use of
the farthest failure position makes it harder to recover from
an error, as the error is only known after parsing finishes and
all the parsing context at the moment of the error has been lost.
Because of this, we will focus on using labeled failures for
error reporting in PEGs.

We need to annotate our original PEG with labels,
which indicate the points where we can signal a
syntactical error.
Figure~\ref{fig:javalabels} annotates the PEG of Figure~\ref{fig:javagrammar} (except for the {\it Prog} rule). 
The expression $[p]^{l}$ is syntactic sugar for $(p \; / \; \throw^{l})$. 
It means that if the matching of $p$ fails we should
throw label $l$ to signal an error.

\begin{figure}[t]
{\small
\begin{align*}
\it{Prog} & \leftarrow {\tt PUBLIC\;CLASS\;NAME\;LCUR\;PUBLIC\;STATIC\;VOID\;MAIN} \\
& \;\;\;\;\;\;\;\;{\tt LPAR\;STRING\;LBRA\;RBRA\;NAME\;RPAR}\;\it{BlockStmt}\;{\tt RCUR}\\
{\it BlockStmt} & \leftarrow {\tt LCUR}\;{\it (Stmt)*}\;[{\tt RCUR}]^{\tt rcblk}\\
{\it Stmt} & \leftarrow {\it  IfStmt\;/\;WhileStmt\;/\; PrintStmt\;/\;DecStmt\;/\;AssignStmt\;/}\\
& \;\;\;\;\;\;\;\;{\it BlockStmt}\\
{\it IfStmt} & \leftarrow {\tt IF\;[LPAR]^{\tt lpif}}\;[{\it Exp}]^{\tt condi}\;[{\tt RPAR}]^{\tt rpif}\;[{\it Stmt}]^{\tt then}\\
& \;\;\;\;\;\;\;\;{\it (}{\tt ELSE}\;[{\it Stmt}]^{\tt else}\;/\;\varepsilon)\\
{\it WhileStmt} & \leftarrow {\tt WHILE\;[LPAR]^{\tt lpw}}\;[{\it Exp}]^{\tt condw}\;[{\tt RPAR}]^{\tt rpw}\;[{\it Stmt}]^{\tt body}\\
{\it DecStmt} & \leftarrow {\tt INT}\;[{\tt NAME}]^{\tt ndec}\;({\tt ASSIGN}\;[{\it Exp}]^{\tt edec}\;/\;\varepsilon)\;[{\tt SEMI}]^{\tt semid}\\
{\it AssignStmt} & \leftarrow {\tt NAME\;[ASSIGN]^{\tt assign}}\;[{\it Exp}]^{\tt rval}\;[{\tt SEMI}]^{\tt semia}\\
{\it PrintStmt} & \leftarrow {\tt PRINT\;[LPAR]^{\tt lpp}}\;[{\it Exp}]^{\tt eprint}\;{\tt [RPAR]^{\tt rpp}\;[SEMI]^{\tt semip}}\\
\it{Exp} & \leftarrow \it{RelExp \; ({\tt EQ} \; [RelExp]^{\tt relexp})*} \\
\it{RelExp} & \leftarrow \it{AddExp \; ({\tt LT} \; [AddExp]^{\tt addexp})*} \\
\it{AddExp} & \leftarrow \it{MulExp \; (({\tt PLUS} \; / \; {\tt MINUS}) \; [MulExp]^{\tt mulexp})*} \\
\it{MulExp} & \leftarrow \it{AtomExp \; (({\tt TIMES} \; / \; {\tt DIV}) \; [AtomExp]^{\tt atomexp})*} \\
\it{AtomExp} & \leftarrow \it{{\tt LPAR} \; [Exp]^{\tt parexp} \; [{\tt RPAR}]^{\tt rpe} \; / \; {\tt NUMBER} \; / \; {\tt NAME}}
\end{align*}
}
\vspace{-0.7cm}
\caption{A PEG with labels for a tiny subset of Java}
\label{fig:javalabels}
\end{figure}

The strategy we used to annotate the grammar was to annotate every
symbol (terminal or non-terminal) in the right-hand side of a production
that should not fail, as failure would just make the whole parser either fail
or not consume the whole input. For an LL(1) grammar, like the
one in our example, that means all symbols in the right-hand side
of a production, except the first one.
We apply the same strategy when the right-hand side has a choice
or a repetition as a subexpression. 

We can associate each label with an error message.
For example, in rule \textit{WhileStmt} the label \texttt{rpw}
is thrown when we fail to match a `\textbf{)}', so we could
attach an error message like ``{\tt missing ')' in while}''
to this label. Dynamically, when the matching of `\textbf{)}'
fails and we throw \texttt{rpw}, we could enhance this
message with information related to the input position
where this error happened.

Let us consider the example Java program from Figure~\ref{fig:javaerror},
which has two syntax errors: a missing `\textbf{)}' at line 5, and a 
missing semicolon at the end of line 7. For this program,
a parser based on the labeled PEG from Figure~\ref{fig:javalabels}
would give us a message like:
\begin{verbatim}
    factorial.java:5: syntax error, missing ')' in while
\end{verbatim}

\begin{figure}[t]
{\small
\begin{verbatim}
    1  public class Example {
    2    public static void main(String[] args) {
    3      int n = 5;
    4      int f = 1;
    5      while(0 < n {
    6        f = f * n;
    7        n = n - 1
    8      }
    9      System.out.println(f);
    10   }
    11 }
\end{verbatim}
}
\caption{A Java program with syntax errors}
\label{fig:javaerror}
\end{figure}

The second error will not be reported because the parser
did not recover from the first one, since {\tt rpw} still has
no recovery expression associated with it.

The recovery expression $p_r$ of an label $l$ matches the input from
the point where $l$ was thrown. If $p_r$ succeeds
then regular parsing is resumed as if the label had not been thrown.
Usually $p_r$ should just skip part of the input until is safe to resume parsing.
In rule \textit{WhileStmt}, we can see that after the
`\textbf{)}' we expect to match a \textit{Stmt}, so the
recovery expression of label \texttt{rpw} could skip the
input until it encounters the beginning of a statement.

In order to define a safe input position to resume parsing,
we will use the classical $\fst$ and $\flw$ sets. 
A detailed discussion about $\fst$ and
$\flw$ sets in the context of PEGs can be found
in other papers~\cite{redz09,redz14,mascarenhas2014}.

With the help of these sets, we can define the following recovery
expression for \texttt{rpw}, where
`\textbf{.}' is a parsing expression that matches any character:
\begin{align*}
(!{\tt \fst(Stmt)} \,.)*
\end{align*}

Now, when label \texttt{rpw} is thrown,
its recovery expression matches the input until
it finds the beginning of a statement, and
then regular parsing resumes. 

The parser will now also throw label {\tt semia}
and report the second error, 
the missing semicolon at the end of line 7.
In case {\tt semia} has an associated recovery
expression, this expression will be used to try
to resume regular parsing again.

Even our toy grammar has 26 distinct labels, each needing
a recovery expression to recover from all possible
syntactic errors. While most of these expressions are
trivial to write, this is still burdensome, and for
real grammars the problem is compounded by the fact that
they can easily need a small multiple of this number of labels.
In the next section, we present an approach to automatically
annotate a grammar with labels and recovery expressions
in order to provide a better starting point for larger grammars.

\section{Automatic Insertion of Labels and Recovery Expressions}
\label{sec:algo}

The use of labeled failures trades better precision in error messages, and
the possibility of having error recovery, for an increased annotation burden,
as the grammar writer is responsible for annotating the grammar with the
appropriate labels. In this section, we show how this process can be
automated for some classes of parsing expression grammars.

To automatically annotate a grammar, we need to determine when it is safe
to signal an error: we should only throw a label after expression $p$ fails
if that failure {\em always} implies that the whole parse will fail
or not consume the whole input, so it is useless to backtrack.

This is easy to determine when we have an $LL(1)$ grammar, as is the case
with the PEG from Figure~\ref{fig:javagrammar}. As we mentioned in
Section~\ref{sec:pegs}, for an $LL(1)$ grammar the general rule is
that we should annotate every symbol (terminal or non-terminal)
in the right-hand side of a production after consuming at least
one token, which in general leads to annotating every symbol 
in the right-hand side of a production except the first one.

Although many PEGs are not $LL(1)$, we can use this approach to
annotate what would be the $LL(1)$ parts of a non-$LL(1)$ grammar.
We will discuss some limitations of this approach in the next section,
when we evaluate its application to the Titan programming language.

After annotating a PEG with labels we can add an automatically
generated recovery expression for each label, based on the tokens
that could follow it.

Algorithm~\ref{alg:annotate}
automatically annotates the parts of a PEG
$G =(V,T,P,L,R,{\tt fail}, p_{S})$. 
We assume that all occurrences of $\fst$
and $\flw$ in Algorithm~\ref{alg:annotate} give
their results regarding to the grammar $G$ passed
to function $\annotate$. We also assume grammar
$G^\prime$ from function $\annotate$ is available
in function $\addlab$.

\begin{algorithm}
{\small
\begin{algorithmic}[1]
\Function{\annotate}{$G$}
	\Let{$G^\prime$}{$G$}	
	\For{$A \in G$}  
		\Let{$G^\prime(A)$}{$\labexp(G(A), \mathbf{false}, \flw(A))$}
	\EndFor
	\State \Return $G^\prime$
\EndFunction

\State

\Function{\labexp}{$p, seq, flw$}
	\If{$p = a \;\,\mathbf{and}\;\, seq$}
		\State \Return $\addlab(p, flw)$
    \ElsIf{$p = A \;\,\mathbf{and}\;\, \Epsi \notin \fst(A) \;\,\mathbf{and}\;\, seq$}
		\State \Return $\addlab(p, flw)$ 
	\ElsIf{$p = p_1\;p_2$}
        \Let{$p_x$}{$\labexp(p_1,seq,\calck(p_2,flw))$}
        \Let{$p_y$}{$\labexp(p_2,seq \;\,\mathbf{or}\;\,  \Epsi \notin \fst(p_1),flw)$}
        \State \Return $p_x \; p_y$
	\ElsIf{$p = p_1 \;/\; p_2$}
        \Let{$p_x$}{$p_1$}
        \If{$\fst(p1) \cap \calck(p2, flw) = \emptyset$}
			\Let{$p_x$}{$\labexp(p_1,\mathbf{false},flw)$}
        \EndIf
        \Let{$p_y$}{$\labexp(p_2,\mathbf{false},flw)$}
		\If{$seq \;\,\mathbf{and}\;\, \Epsi \notin \fst(p_1 \;/\; p_2)$}
			\State \Return $\addlab(p_x \;/\; p_y, flw)$ 
		\Else
			\State \Return $p_x \;/\; p_y$
		\EndIf
	\ElsIf{$p = p_1\!* \;\,\mathbf{and}\;\, \fst(p_1) \cap flw = \emptyset$}
		\State \Return $\labexp(p_1, \mathbf{false}, flw)*$
	\Else
		\State \Return $p$
	\EndIf
\EndFunction

\State

\Function{\calck}{$p,flw$}
	\If{$\Epsi \in \fst(p)$}
	    \State \Return ($\fst(p) - \{ \Epsi \}) \cup flw$
	\Else  
		\State \Return $\fst(p)$
	\EndIf
\EndFunction

\State

\Function{\addlab}{$p, flw$}
  \Let{$l$}{newLabel()} 
  \Let{$\Rp(l)$}{$(!flw \; .)*$}
	\State \Return $[p]^l$  
\EndFunction
\end{algorithmic}
}
\caption{Automatically Inserting Labels and Recovery Expressions in a PEG}
\label{alg:annotate}
\end{algorithm}

Function \texttt{\annotate} (lines 1--5) generates a new annotated grammar $G^\prime$
from a grammar $G$. It uses \texttt{\labexp} (lines 7--28) to annotate the
right-hand side, a parsing expression, of each rule of grammar G. The
auxiliary function \texttt{\calck} (lines 30--34) is used to update the
$\flw$ set associated with a parsing expression. By its turn, the auxiliary
function \texttt{\addlab} (lines 36--39) receives a parsing expression $p$ to annotate
and its associated $\flw$ set $flw$. Function \texttt{\addlab} associates a label $l$ to $p$ and also
builds a recovery expression for $l$ based on $flw$.

We annotate every right-hand side, instead of going top-down from the root,
to not be overly conservative and fail to annotate non-terminals
reachable only from non-LL(1) choices but which themselves might be LL(1).
We will see in Section~\ref{sec:eval} that this has the unfortunate result
of sometimes changing the language being parsed, and is the major
shortcoming of our algorithm.

Function \texttt{\labexp} has three parameters. The first one, $p$, is
a parsing expression that we will try to annotate. The second parameter,
$seq$, indicates whether we have already matched a prefix of a concatenation 
expression that consumes at least one terminal or not. Parameter $seq$ has
value {\bf true} when $p$ is a suffix of a concatenation $p_1\; p_2$
and the prefix of $p_1\;p_2$ already consumed at least one input character.
Finally, the parameter $flw$ represents the $\flw$ set associated with $p$.
Let us now discuss how \texttt{\labexp} tries to annotate $p$.

When $p$ is an expression that matches a terminal and 
is part of a concatenation that already matched
at least one terminal (lines 8--9), then we associate a new
label with $p$. In case $p$ represents a terminal but $seq$
is not \textbf{true}, we will just return $p$ (lines 27--28).

The case when $p$ matches a non-terminal $A$ is similar (lines 10--11),
we have just added an extra condition that tests whether $A$ matches
the empty string or not. This avoids polluting the grammar
with labels which will never be thrown, since a parsing expression
that matches the empty string does not fail.

In case of a concatenation $p_1\;p_2$ (lines 12--15), we try to annotate
$p_1$ and $p_2$ recursively. To annotate $p_1$ we use an updated $\flw$ set, 
and to annotate $p_2$ we set its parameter $seq$ to \textbf{true} whenever
$seq$ is already \textbf{true} or $p_1$ does not match the empty string.

In case of a choice $p_1\,/\,p_2$ (lines 16--24), we annotate
$p_2$ recursively and in case the choice is disjoint we also
annotate $p_1$ recursively. In both cases, we pass the value
\textbf{false} as the second parameter of \texttt{\labexp},
since failing to match the first symbol of an alternative should
not signal an error. When $seq$ is \textbf{true}, we associate
a label to the whole choice when it does not match the empty
string. 

In case $p$ is a repetition $p_1*$ (lines 25--26), we can annotate $p_1$
if there is no intersection between $\fst(p_1)$ and $flw$. 
When annotating $p_1$ we pass \textbf{false} as the second parameter
of \texttt{\labexp} because failing to match the first symbol of a
repetition should not signal an error.

Given the PEG from Figure~\ref{fig:javagrammar}, function \texttt{\annotate}
would give us the grammar presented in Figure~\ref{fig:javalabels}
(as previously, we are not taking rule {\it Prog} into consideration),
with the exception of the annotation $[{\it Stmt}]^{\tt else}$.
Label {\tt else} was not inserted at this point because token
{\tt ELSE} may follow the choice ${\tt ELSE}\,{\it Stmt\;/\;\Epsi}$,
so this choice is not disjoint (the well-known {\it dangling else} problem).

It is trivial to change the algorithm to leave any existing labels
and recovery expressions in place, or to add recovery expressions to
any labels that are already present but do not have recovery expressions.

After applying Algorithm~\ref{alg:annotate} to automatically
insert labels, a grammar writer can later add (or remove) labels and
their associated recovery expressions.
We discuss more about this on the next section, where we evaluate
the use of Algorithm~\ref{alg:annotate} to add error recovery for
the Titan programming language.

\section{Evaluation}
\label{sec:eval}

Titan~\cite{titan} is a new statically-typed programming language
under development to be used as a sister language to the Lua programming
language~\cite{lua}.

The Titan parser uses LPegLabel~\cite{lpeglabel},
a tool that implements the semantics of PEGs with labeled failures,
and has labels that were manually inserted. The syntactical rules of Titan
parser can be seen at the following link:
\url{https://goo.gl/Thqi9r}.

The 50 syntactical rules of Titan grammar have around
85 expressions that throw labels. Some labels, such as {\tt EndFunc},
are thrown more than once.
The grammar has no error recovery, and aborts the parser on encountering the
first syntax error.

The Titan grammar is not $LL(1)$, but has many $LL(1)$ parts, 
and the Titan developers intend to keep using a PEG-based parser 
for the Titan compiler, so it seemed
a good candidate to evaluate our algorithm that automatically insert
labels and its corresponding recovery expressions. 

In order to do this, we first wrote an unlabeled version of the Titan
grammar, without all labels and some lexical elements, but with the same
syntactical rules of the original grammar\footnote{This grammar
is available at 
\url{https://github.com/sqmedeiros/recoveryPEG/blob/master/gramTitan.lua}.}.

We then wrote a program to apply Algorithm~\ref{alg:annotate} to
this unlabeled grammar and got an automatically annotated Titan grammar,
with a recovery expression associated to each label\footnote{This grammar
is available at
\url{https://github.com/titan-lang/titan/blob/recovery/titan-compiler/annotatedTitan.txt}.}. 
Throughout this section we will use the term {\it generated} to
refer to this grammar.

Our first evaluation, in Section~\ref{sec:autolab}, compares
the labels automatically inserted with the labels in the original
Titan grammar. Then, in Section~\ref{sec:autorec}, we will discuss the error
recovery mechanism of the generated Titan grammar.

\subsection{Automatic Insertion of Labels}
\label{sec:autolab}

Algorithm~\ref{alg:annotate} annotates the Titan
grammar with 79 labels, which is close to the number of labels of the
original Titan grammar. A manual inspection revealed that the algorithm
inserted labels at the same location of the original ones, although there
are some caveats that we should talk about. Below we will discuss the
following points:
\begin{enumerate}
    \item \label{it:notlab} When the algorithm did not insert a label.
    \item \label{it:wronglab} When the algorithm incorrectly inserted a label.
    \item \label{it:newlab} When the algorithm correctly inserted a new label. 
\end{enumerate}

About Item~\ref{it:notlab}, as expected our approach did not annotate parts of the 
grammar where the alternatives of a choice were not disjoint. This happened
in 4 of the 50 grammar rules. One of these rules was \textit{castexp},
which we show below:
\begin{align*}
{\it castexp} & \leftarrow {\it simpleexp} \;{\tt AS}\; {\it type} \;/\; {\it simpleexp}
\end{align*}

As we can see, both alternatives of the choice match a {\it simpleexp},
so these alternatives are not disjoint. After manual inspection, we can
see it is possible to add a label to {\it type} in the first alternative, 
since the context where {\it castexp} appears in the rest of the grammar
makes it clear that a failure on {\it type} is always a syntax error.
Left-factoring the right-hand side of {\it castexp} to ${\it simpleexp}\;({\tt 
AS}\;{\it type}\;/\;\varepsilon)$ would give enough context for Algorithm~\ref{alg:annotate}
to correctly annotate {\it type} with a label, though.

The original Titan grammar also uses an approach known
as {\it error productions}~\cite{grune2010ptp}.
As an example, the choice associated with rule
{\it statement} has two extra alternatives whose only purpose
it to match some usual syntactically invalid statements, in order
to provide a better error message.
One of these alternatives is as follows:
\begin{align*}
    \&({\it exp}\; ASSIGN)\; \throw^{{\tt ExpAssign}}
\end{align*}

Before this alternative, the grammar has one that tries to
match an assignment statement. That alternative might have
failed because the programmer used an expression that is not
a valid l-value in the left-hand side of the assignment.
This error production guards against this case. Without the error production, 
the next alternative, which tries to recognize a statement that is a single {\it call} expression, would report an inappropriate error.
When given an input like `\texttt{3 = x}', the next alternative
would match `{\tt 3}' and then we would get an unrelated error when trying to match the assignment operator.

Again, this could have been worked around if the grammar had been changed slightly by, for example, adding a
$!{\tt ASSIGN}$
predicate at the end of the alternative that matches
expression statements.

To a certain extent, we do not consider not adding labels as a serious issue, 
as long as most of the other labels are correctly inserted, since 
failing to add labels does not lead to an incorrect parser. These
(hopefully few) labels can still be manually inserted later by an expert.

A discrepancy related to Item~\ref{it:wronglab} is more problematic,
since it can produce a parser that does not recognize some
syntactically valid programs. This issue happened in rules {\it toplevelvar}
and {\it import}. Figure~\ref{fig:titanimport} shows the definition of these
rules, plus some rules that help to add context, in the original Titan grammar.
Lexical elements are show in teletype and uppercase, keeping the typographical style we
have been using.

\begin{figure*}
{\small
\begin{align*}
{\it program} & \leftarrow  ( {\it toplevelfunc} \;/\; {\it toplevelvar} \;/\; {\it toplevelrecord} \;/\; {\it import} \;/\; {\it foreign} )* \;\,!{\tt .}\\
{\it toplevelvar} & \leftarrow {\it localopt}\; {\it decl}\; [{\tt ASSIGN}]^{\tt AssignVar}\; !({\tt IMPORT} \;/\; {\tt FOREIGN})\; [{\it exp}]^{\tt ExpVarDec}  \\
{\it import}      & \leftarrow  {\tt LOCAL}\, [{\tt NAME}]^{\tt NameImport}\; [{\tt ASSIGN}]^{\tt AssignImport} \;
                          !{\tt FOREIGN}\; [{\tt IMPORT}]^{\tt ImportImport}\;
                          (\,{\tt LPAR}\; [{\tt STRING}]^{\tt StringLParImport}\; [{\tt RPAR}]^{\tt RParImport} \;/\;
                          [{\tt STRING}]^{\tt StringImport}) \\
{\it foreign}      & \leftarrow  {\tt LOCAL}\, [{\tt NAME}]^{\tt NameImport}\, [{\tt ASSIGN}]^{\tt AssignImport} \,
                          {\tt FOREIGN} \, [{\tt IMPORT}]^{\tt ImportImport}\,
                          (\,{\tt LPAR}\, [{\tt STRING}]^{\tt StringLParImport}\, [{\tt RPAR}]^{\tt RParImport} \,/\,
                          [{\tt STRING}]^{\tt StringImport}) \\
{\it decl} & \leftarrow {\tt NAME}\; (\,{\tt COLON}\; [{\it type}]^{\tt TypeDecl})?  
\end{align*}
}
\caption{Predicates $!({\tt IMPORT} / {\tt FOREIGN})$ and $!{\tt FOREIGN}$ enable the insertion of labels after them} 
\label{fig:titanimport}
\end{figure*}

Rules {\it toplevelvar}, {\it import} and {\it foreign} are alternatives of a choice in rule {\it program}. The parser first tries to recognize
{\it toplevelvar}, then {\it import}, and finally {\it foreign}.
As a {\it decl} may consist of only a name, an input like
`\texttt{local x =}' may be the beginning of any of these rules.
In rule {\it toplevelvar}, the predicate $!({\tt IMPORT} / {\tt FOREIGN})$
was added by the Titan developers to make sure the input does not match the {\it import} or
the {\it foreign} rules, so it is safe to throw an error after this
predicate in case we do not recognize an expression. The predicate $!{\tt FOREIGN}$
in rule {\it import} plays a similar role. 

These predicates were not present in our {\it unlabeled} grammar\footnote{These predicates were inserted by
the Titan authors solely to enable the subsequent
label annotations, so we judged that it was a fairer
evaluation to also remove them from our unlabeled
grammar.}.
Alternatives {\it toplevelvar}, {\it import}, and {\it foreign} all have
{\tt LOCAL} in their $\fst$ sets, but the algorithm adds labels to the right-hand
side of these non-terminals under the assumption that they are not used on a
non-LL(1) choice.

The outcome is that the algorithm is able to insert the same labels as the
original grammar, but without the syntactic predicates these labels wrongfully
signal errors in valid inputs such as `\texttt{local x = import "foo"}'.

This limitation of our algorithm means that the output needs to be checked
by the parser developer, to make sure that non-LL(1) parts of the grammar
have not been broken. We still believe this is less work than manually
annotating the grammar, given that the parser already needs to have an
extensive test suite that will catch these errors, as was the case in
our evaluation.

For the generated Titan grammar, this problem only happened
in rules {\it toplevelvar} and {\it import}. We fixed the issue by adding
the corresponding predicates, which were already present in the original annotated
grammar.

Finally, Item~\ref{it:newlab} represents a situation where our algorithm
was able to insert a label that was not inserted manually, although it
should be. This happened once, in rule {\it type}, where the algorithm
annotated `{\bf ->}' in the first alternative.

After fixing the issues related to Item~\ref{it:wronglab}, our
generated Titan parser successfully passed the Titan tests.
For completeness, we also manually added the missing labels
to the generated grammar. 

\subsection{Automatic Error Recovery}
\label{sec:autorec}

When inserting labels, the Algorithm~\ref{alg:annotate} also
associates a recovery expression to each label.

In LPegLabel, to recover from a label $l$ we need to add
a recovery rule $l$ to our grammar, where the right-hand side of
$l$ is its recovery expression. Our generated Titan grammar has
a recovery rule associated with each label.

LPegLabel does not try to automatically generate an AST for
grammar rules, so we still needed to manually add semantic
actions to generate valid AST nodes in case of error recovery.
A more high-level tool, in the spirit of
ANTLR~\cite{parr2013antlr}, could infer these
semantic actions automatically, based on the type of
the AST node produced at the point of the error.

For our evaluation of automatic error recovery we took the original
Titan parser and added the extra label that our algorithm inferred plus
the recovery expressions for the other labels, along with manually
annotated semantic actions for these recovery expressions.

We again ran the test set of Titan to see whether our parser
with error recovery would recover from all syntax errors
and build valid ASTs for syntactically invalid programs.

The test suite of Titan had 77 tests related to syntactically
invalid programs. For all these tests, with the exception of one,
our error recovery strategy could build an AST. For 70 of these tests
the recovery strategy resulted in just one syntax error. 

The input where our recovery strategy failed was the following one:
\begin{verbatim}
local x: (a, b) -> = nil = nil
\end{verbatim}

In this case, a correct program would be something as follows,
where there is a \texttt{nil} after the `{\tt ->}':
\begin{verbatim}
local x: (a, b) -> nil
\end{verbatim}

Given the previous invalid input, our parser throws an error when
it finds the first `{\tt =}'
and correctly recovers from it, producing a dummy AST node
for the missing return type. The problem happens when the
parser finds the second `{\tt =}', that is not part of {\tt x}'s declaration
and can not be the start of any alternative in rule {\it program}
of Titan grammar (see this rule in Figure~\ref{fig:titanimport}).
Thus, the repetition in rule {\it program} fails and the end of
input is expected, but as the input suffix `{\tt = nil}' was not
matched we get an error. 

In this particular example we could have just thrown away the second
`{\tt = nil}' in the input, but in general we could have be throwing away
a significant (and correct) part of the input.

We could have adapted Algorithm~\ref{alg:annotate} to deal
with this case by changing the case of $p_1^*$ in function
\addlab as follows: 
\begin{align*}
(!flw \; \addlab(\labexp(p_1, \mathbf{false}, flw)))*
\end{align*}

This way, whenever we fail to match $p_1$, as we know
that the matching of $flw$ will fail, we throw a label.
After the recovery expression associated with this label
synchronizes with the input, we will try to match $p_1$
again, whether still is not possible to match its follow
expression. 

The problem with applying this approach for every
parsing expression such as $p*$, $p?$, and $p+$
in our grammar is that we would add too many labels, going from
80 labels to about 120, and these labels would be too specific.
The above example would log an error in the binary expression
rule with the greatest precedence level in the Titan grammar, as
this rule has a repetition where the second
`{\tt = nil}' is not in its $\flw$ set. This error message
would not be informative to the programmer.
Moreover, the more labels we add the greater the possibility
for introducing bugs in the grammar related to Item~\ref{it:wronglab} above.

Having said that, we think it would be useful to label $p*$
when $p$ matches a high level rule of the grammar, as it
is the case of the repetition in rule {\it program} of
Titan (see Figure~\ref{fig:titanimport}), and in rule
{\it BlockStmt} of our Java grammar (see Figure~\ref{fig:javalabels}).
A grammar does not have many of these rules, though, so
it is not onerous to require manual annotations in these cases.

After discussing the one case where our error recovery
approach failed, let us now see the other 76 cases.
The fact that applying our recovery 
resulted in just one syntax error for almost $90\%$
of the syntactically invalid programs seems  
to indicate that the recovery expression
we used helped to resume parsing at an appropriate input position.
However, to better evaluate our recovery strategy we still need to
check how much information about a syntactically invalid program our
AST usually has.

Pennelo and DeRemmer~\cite{pennello1978forward} proposed
evaluating the quality of an error recovery approach based
on the similarity of the program obtained after recovery with the
intended program (without syntax errors). This quality measure
was used to evaluate several
strategies~\cite{corchuelo2002repair,degano1995comparison,dejonge2012natural},
although it is arguably subjective~\cite{dejonge2012natural}.
 
We will evaluate our strategy following Pennelo and DeRemmer's approach, 
however instead of comparing programming texts we will compare the AST 
from an erroneous program after recovery
with the AST of what would be an equivalent correct program.
For the leaves of our AST we do not require their contents to be
the same, just the general type of the node, so we are comparing
just the structure of the ASTs.

Based on this strategy, a recovery is {\it excellent}
when it gives us an AST equal to the intended one.
A {\it good} recovery gives us a reasonable AST, i.e.,
one that captures most information of the original program
(e.g., it does not miss a whole block of commands),
does not report spurious errors, and does not miss other errors.
A {\it poor} recovery, by its turn, produces an AST that
loses too much information, results in spurious errors, or
misses errors. Finally, a recovery is rated as {\it failed}
whenever it fails to produce an AST at all.

Table~\ref{tab:eval1} shows for how many programs the recovery
strategy we implemented was considered {\it excellent},
{\it good}, {\it poor}, or {\it failed}. As we can see,
our recovery mechanism for Titan seems promising, since that
more than 90\% of the recovery done was considered acceptable,
i.e., it was rated at least {\it good}.

\begin{table}[t]
    \centering
    \begin{tabular}{|c|c|c|c|c|}
      \hline
      Excellent &            Good & Poor & Failed  \\ \hline
      68 ($\approx 88\%$) & 4 ($\approx 5\%$)   & 4 ($\approx 5\%$)   & 1 ($\approx 1\%$)       \\
    \hline
    \end{tabular}
\vspace{0.5cm}
    \caption{Evaluation of an automatic error recovery in the Titan Parser}
    \label{tab:eval1}
\vspace{-0.5cm}
\end{table}

As most of programs of Titan test suite are small,
we plan later to evaluate our recovery mechanism
by using bigger test programs.

\section{Related Work}
\label{sec:rel}

In this section, we discuss some error reporting and recovery approaches
described in the literature or implemented by parser generators. 

Swierstra and Duponcheel~\cite{swierstra1996dec} show an
implementation of parser combinators for error recovery,
but it is restricted to LL(1) grammars. The recovery strategy
is based on a {\em noskip} set, computed by taking the
$\fst$ set of every symbol in the tails of the pending
rules in the parser stack. Associated with each token in this set
is a sequence of symbols (including non-terminals) that would
have to be inserted to reach that point in the parse, taken from
the tails of the pending rules. Tokens are then skipped until
reaching a token in this set, and the parser then takes actions
as if it have found the sequence of inserted symbols for this token.

Our approach cannot simulate this recovery strategy, as it relies
on the path that the parser dynamically took to reach the point of
the error, while our recovery expressions are statically determined
from the label. But while their strategy is more resistant to
the introduction of spurious errors than just using the $\flw$
set it still can introduce those.

A popular error reporting approach applied for bottom-up parsing
is based on associating an error message to a parse state and a
lookahead token~\cite{jeffery2003lr}. To determine the error
associated to a parse state, it is necessary first to manually
provide a sequence of tokens that lead the parser to that failure state.
We can simulate this technique with the use of labels. By using
labels we do not need to provide a sample invalid program for
each label, but we need to annotate the grammar properly.

Coco/R~\cite{cocomanual} is a tool that generates predictive
$LL(k)$ parsers. As the parsers based on Coco/R do not backtrack,
an error is signaled whenever a failure occurs. In case of PEGs,
as a failure may not indicate an error, but the need to backtrack,
in our approach we need to annotate a grammar with labels, a task
we tried to make more automatic.

In Coco/R, in case of an error the parser reports it and
continues until reaching a {\it synchronization point}, which
can be specified in the grammar by the user through the use of a keyword
{\tt SYNC}. Usually, the beginning of a statement or
a semicolon are good synchronization points. 

Another complementary mechanism used by Coco/R for error
recovery is {\tt weak} tokens, which can be defined by
a user though the {\tt WEAK} keyword. A weak token is one that
is often mistyped or missing, as a comma in a parameter list,
which is frequently mistyped as a semicolon. When  the parser fails
to recognize a weak token, it tries to resume parsing
based also on tokens that can follow the weak one.

Labeled failures plus recovery expressions can simulate the
{\tt SYNC} and {\tt WEAK} keywords of Coco/R. 
Each use of {\tt SYNC} keyword would correspond to
a recovery expression that advances the input 
to that point, and this recovery expression would
be used for all labels in the parsing extent of this
synchronization point.
A {\tt WEAK} can have a recovery expression that
tries also to synchronize on its $\flw$ set.

Coco/R avoids spurious error messages during
synchronization by only reporting an error if at least two tokens
have been recognized correctly since the last error. This is easily
done in labeled PEG parsers through a separate post-processing step.

ANTLR~\cite{antlrsite,parr2013antlr} is a popular tool
for generating top-down parsers. 
ANTLR automatically generates from a grammar description a parser
with error reporting and
recovery mechanisms, so the user does not need to annotate
the grammar. After an error, ANTLR parses the
entire input again to determine the error,
which can lead to a poor performance when
compared to our approach~\cite{medeiros2018sac}.

As its default recovery strategy, ANTLR attempts single
token insertion and deletion to synchronize with the input. In case the 
remaining input can not be matched by any production of the 
current non-terminal, the parser consumes the input 
``\textit{until it finds a token that could reasonably follow
the current non-terminal}''~\cite{parr2014antlr}.
ANTLR allows to modify the default error recovery approach,
however, it does not seem to encourage the definition of a 
recovery strategy for a particular error,
the same recovery approach is commonly used for the whole
grammar.

A common way to implement error recovery in PEG parsers
is to add an alternative to a failing expression,
where this new alternative works as a fallback. Semantic actions
are used for logging the error.
This strategy is mentioned in the manual of Mouse~\cite{redzmouse}
and also by users of LPeg\footnote{See 
\url{http://lua-users.org/lists/lua-l/2008-09/msg00424.html}}.
These fallback expressions with semantic actions for error logging
are similar to our recovery expressions and labels, but in an ad-hoc,
implementation-specific way.

Several PEG implementations such as
Parboiled\footnote{\url{https://github.com/sirthias/parboiled/wiki}},
Tatsu\footnote{\url{https://tatsu.readthedocs.io}},
and PEGTL\footnote{\url{https://github.com/taocpp/PEGTL}} provide 
features that facilitate error recovery.

The previous version of Parboiled used an error recovery
strategy based on ANTLR's one, and requires parsing
the input two or three times in case of an error.
Similar to ANTLR, the strategy used by Parboiled 
was fully automated, and required neither manual intervention
nor annotations in the grammar. Unlike ANTLR, it
was not possible to modify the default error strategy.
The current version of
Parboiled\footnote{\url{https://github.com/sirthias/parboiled2/}}
does not has an error recovery mechanism.

Tatsu uses the fallback alternative technique for error
recovery, with the addition of a {\it skip expression},
which is a syntactic sugar for defining a pattern that
consumes the input until the skip expression succeeds. 
PEGTL allows to define for each rule $R$ a set of
terminator tokens $T$, so when the matching of $R$
fails, the input is consumed until a token $t \in T$
is matched. This is also similar to our approach for recovery
expressions, but with coarser granularity, and lesser control
on what can be done after an error.

Rüfenacht~\cite{michael2016error} proposes a local
error handling strategy for PEGs. This strategy uses
the farthest failure position and a record of the parser
state to identify an error. Based on the information
about an error, an appropriate recovery set is used.
This set is formed by parsing expressions that match
the input at or after the error location, and it is used
to determine how to repair the input. 

The approach proposed by Rüfenacht is also similar to the use
of a recovery expression after an error, but more limited
in the kind of recovery that it can do. When testing his approach in
the context of a JSON grammar, which is simpler than the Titan grammar,
Rüfenacht noticed long running test cases and mentions the need to
improve memory use and other performance issues.

\section{Conclusion}
\label{sec:conc}

We have presented a mechanism for partially automating the process
of adding error recovery to parsers based on Parsing Expression Grammars,
using an algorithm that automatically annotates the LL(1) parts of
a PEG with error labels~\cite{maidl2013peglabel,maidl2016peglabel} and recovery
expressions for these labels~\cite{medeiros2018sac}. 

The parser developer still has to check whether there are parts
of the grammar that should be annotated but were not. In case
generic error messages that only indicate which term was expected
and what was found in the input are not enough, the parser developer
also needs to associate specific error messages with each inserted label.

We evaluated our algorithm on the parser for the Titan programming
language, which had already been heavily annotated by its developers,
and has an extensive test suite both for valid and erroneous input.
For our evaluation, we removed all labels from the Titan grammar,
and compared the output of our algorithm with the original annotations.
Our algorithm reinserted 78 of the original 85 annotations, plus one
annotation that was not in the original grammar. Our automatically-generated
recovery expressions lead to excellent error recovery in 88\% of the
cases in the Titan test suite where there are invalid inputs, and
only fails to generate an AST for 1\%.

While our evaluation has started by removing annotations on a fully-annotated
grammar, the algorithm is easily extended to work on grammars that have
already been partially annotated, either with just labels or labels and
recovery expressions, as well as marking the parts of the grammar where the
algorithm had to ignore and should be annotated
by hand by the parser developer.

The major limitation of our algorithm is the LL(1) assumption it makes,
specially the non-conservative part where it annotates the right-hand side
of every non-terminal as if it were in the LL(1) part of the grammar, which
can lead to insertion of labels that introduce bugs in the grammar (change
its underlying language). We are working on increasing the precision of
the analysis to not introduce spurious annotations, while not decreasing
the number of useful annotations our algorithm inserts. We should also
investigate whether the use of some normal form when writing a PEG grammar
could help our algorithm to produce a better result, without imposing
too much restrictions for a grammar writer.

As another future work, we intend to apply our algorithm to more complex
grammars, such as the grammars of Java and C.

\bibliographystyle{ACM-Reference-Format}
\bibliography{sblp2018}

\end{document}